\documentstyle[11pt,epsfig,color]{article}
\begin{document}

\begin{center}
{\Large\bf Towards a characterization of fields leading to black hole hair}
\\[15mm]
Narayan Banerjee, \footnote{E-mail: narayan@iiserkol.ac.in}\\
{\em Department of Physical Sciences,~~\\Indian Institute of Science and Educational Reserch, Kolkata,~~\\Mohanpur Campus, West Bengal 741252, India}\\
and\\
Somasri Sen \footnote{E-mail: ssen@jmi.ac.in}\\ 
{\em Department of Physics, Jamia Millia Islamia, Jamia Nagar, ~~\\New Delhi 110 025, India}

\end{center}

\vspace{0.5cm}
\vspace{0.5cm}

\pagestyle{myheadings}
\newcommand{\be}{\begin{equation}}
\newcommand{\ee}{\end{equation}}
\newcommand{\bea}{\begin{eqnarray}}
\newcommand{\eea}{\end{eqnarray}}

\begin{abstract}
In the present work, it is shown that an asymptotically flat spherical black hole can have a nontrivial signature of any field for an exterior observer if the energy momentum tensor of the corresponding field is either tracefree or if the trace falls off at least as rapidly as inverse cube of the radial distance. In the absence of a general No Hair Theorem, this result can provide a characterization of the fields leading to black hole hair.
\end{abstract}

\section{Introduction}
Black holes are easily amongst the most fascinating offshoots of General Theory of Relativity. One important question asked about a black hole is that which informations one can extract from the exterior gravitational field of such an object. The answer is normally given in terms of a ``No Hair Theorem''. The so-called theorem says that no information regarding a black hole can be obtained by an exterior observer except that of the mass ($M$), electric charge ($Q$) and the angular momentum ($h$).  This was summarized by Ruffini and  Wheeler\cite{wheeler}. The idea was inspired by the uniqueness of Scwarzchild and Reissner-Nordstrom solutions as shown by Israel\cite{israel} and the uniqueness of Kerr Black hole by Wald\cite{wald} and Carter\cite{carter}. Although the statement proved to be extremely powerful, it should perhaps be called a No Hair Conjecture rather than a theorem as there is hardly any rigorous proof for the same. Attempts are normally made to find examples either in support or to the contrary of the conjecture.  
\par There have been excellent attempts towards a proof of the No Hair Theorem but they normally include a particular field at one go. For example, recently Bhattacharya and Lahiri\cite{lahiri1} proved the theorem for an axially symmetric black in case of a scalar field or a massive vector field. A proof for a No Hair Theorem for a spherical black hole regarding Higgs model is also available in the literature\cite{lahiri}.
\par 
The search for the possibility of information regarding a particular field, i.e., a possible ``hair'', started way back in the early seventies\cite{chase}. The quest gave rise to a classification of black hole hair into two categories, namely a primary hair and a secondary hair\cite{coleman}. A primary hair is one which is independemt of the existence of any other hair. A secondary hair, on the other hand, depends on the existence of other fields and grows on them. The electric field, for example, is a primary hair. The recently discovered dilaton hair\cite{garfinkle} in fact grows on the electric charge and ceases to exist if the electric field is switched off. This is an example of a secondary hair.
\par Indeed there are examples of black hole solutions with a hair other than $M$, $Q$ or $h$, which contradict the No Hair Conjecture. But all these counter examples have some pathology or unwanted features, particularly if the example is that of a primary hair. Most of the black hole solutions with such a hair are unstable\cite{zhou}. The most talked about counter example is the existence of a scalar hair for a conformally invariant scalar field nonminimally coupled to gravity given by Bekenstein\cite{beken}. In this example, the effective Newtonian constant of gravity may become negative! However, for a nonminimally coupled scalar-tensor theory this possibility may not be ruled out. Some axisymmteric black hole solutions endowed with a scalar hair have been found out\cite{ssnb}. These black holes are, however, not asymptotically flat. There are also examples of the possibility of some hair for tiny black holes (i.e., not of the size of a stellar black hole) as discussed by Weinberg\cite{weinberg}.\\
\par Anyway, only this kind of examples are there in the literature regarding the existence of a blach hole hair. Except for the attampts with specific fields\cite{lahiri1, lahiri}, there is hardly any proof or, for that matter, a definite way to characterize the matter fields which may lead to a hair.
\par In the present work, an attempt is made towards the characterization of matter fields which might be detected by an exterior observer. It is shown that for a particular hair to exist for an asymptotically flat spherical black hole, the energy-momentum tensor for the corresponding field must either be trace-free or the trace should fall off with the proper radius $r$ at least as fast as $\frac{1}{r^{3}}$. The scope of this result is a bit limited, as it is achieved only for a spherical black hole, but it definitely gives some indications in the absence of a more rigorous theorem.

\section{The Theorem}
A static spherically symmetric line element has the form
\be
ds^{2}=e^{\nu}dt^{2}-e^{\lambda}dr^{2}-r^{2}(d{\theta}^{2}+sin^{2}{\theta}d{\phi}^{2}),
\ee
where $\nu$ and $\lambda$ are functions of $r$ alone. This so-called curvature form of the metric has the advantage that the radial coordinate $r$ has the significance of the proper radial distance.
\par For this metric, Einstein field equations,
\be
G^{\alpha}_{\beta} = R^{\alpha}_{\beta} - \frac{1}{2} R{\delta}^{\alpha}_{\beta} = -8\pi G T^{\alpha}_{\beta}, 
\ee 
are written as,
\be
e^{-\lambda}\big(\frac{1}{r^{2}} - \frac{{\lambda}^{\prime}}{r}\big) -\frac{1}{r^{2}} = - 8\pi G T^{0}_{0},
\ee
\be
e^{-\lambda}\big( \frac{{\nu}^{\prime}}{r} + \frac{1}{r^{2}}\big) - \frac{1}{r^{2}}= - 8\pi G T^{1}_{1},
\ee
and,
\be
\frac{1}{2} e^{-\lambda}\big( {\nu}^{\prime\prime} +\frac{1}{2} {\nu}^{\prime 2} + \frac{{\nu}^{\prime} - {\lambda}^{\prime}}{r} -\frac{1}{2} {\nu}^{\prime}{\mu}^{\prime}\big) = - 8\pi G T^{2}_{2} = - 8\pi G T^{3}_{3},
\ee
where a prime indicates a differentiation with respect to $r$.
\par The existence of an event horizon characterizes a black hole. The event horizon is a null surface where $g^{11}=0$ ( for a diagonal metric which would also mean $g_{11}$ is infinity). However, the event horizon should be a regular surface which does not have any singularity on the surface. So this apparent singularity of the metric should be an artifact of choice of coordinates and the physical quantities should be well behaved. For example, the curvature should be finite, the proper volume (given by $\sqrt{-g}$) should also be finite and nonzero.
\par With a generalized coordinate condition, given by Duan et al\cite{duan}, the metric can in fact be written in spherical polar coordinates in such a way that 
\be
g_{00}g_{11}=-(\frac{dF}{dr})^{2}
\ee 
where $F$ is the proper radius. As $r$ is the proper radius in the present form of the metric, this condition yields $e^{\nu + \lambda} = 1$. We shall utilize this without any serious loss of generality. This also ensures that $\sqrt{-g}$ is nonzero on the horizon. Schwarzchild solution indeed has this property.
\par A contraction of equation (2) yields 
\be
R = R^{\alpha}_{\alpha} = 8\pi G T^{\alpha}_{\alpha} = 8\pi G T.
\ee
From the expressions for $G^{\alpha}_{\beta}$ given in the left hand side of the field equations (3) to (5), one can write the Ricci scalar $R$ as 
\be
R = \frac{(e^{-\lambda} r^{2})^{\prime\prime}}{r^{2}} -\frac{2}{r^{2}},
\ee
where the condition $e^{\nu + \lambda} = 1$ has been used. This equation, on integration, yields
\be
g_{00} = e^{-\lambda} = 1 + \frac{C_{1}}{r} + \frac{C_{2}}{r^{2}} + \frac{1}{r^{2}} \int (\int R r^{2} dr)dr,
\ee
$C_{1}$, $C_{2}$ being constants of integration. An event horizon is given by a null surface which requires
\be
- g^{11} = e^{-\lambda} = 1 + \frac{C_{1}}{r} + \frac{C_{2}}{r^{2}} + \frac{1}{r^{2}} \int (\int R r^{2} dr)dr = 0,
\ee
The real solutions (of $r$) for this equation will locate the site of event horizons. The number of possible horizons will depend on the degree of the algebraic equation (10) in $r$.
\par If the spacetime is asymptotically flat, one has $g_{00} \sim 1$ when $r$ goes to  $\infty$. So the last term on the right hand side of equation (9) should either be zero or be such that it goes to zero as $r$ approaches $\infty$. This feature is achieved if $R$ falls off as $\frac{1}{r^{3}}$ or faster, for large values of $r$.
\par To facilitate this idea, we now assume that $R$ can be written as a series expansion of $r$ as 
\be
R = {\Sigma}_{i} a_{i}r^{i} + {\Sigma}_{j}b_{j}r^{-j},
\ee
where $a_{i}, b_{j}$ are constants, $i$ runs from $0$ to $m$ and $j$ runs from $0$ to $n$. It is easy to see that if asymptotic flatness is invoked, i.e., $g_{00} \sim 1$  for $r$ approaching $\infty$, one would require to have only inverse powers of $r$ in the expression for $g_{00}$, which, in turn would require that $a_{i}$'s are all zero. So the Taylor series part does not contribute. From the Laurent series part, one can now evaluate the integral in the equation (9) as
\be
\int (\int R r^{2} dr)dr = {\Sigma}_{j}{\alpha}_{j} r^{4-j},
\ee
${\alpha}_{j}$'s are constants. It is now evident that the last term of equation (10) will satisfy the requiremet if $j>3$, i.e., all ${\alpha}_{j} = 0$ for $j<3$. However, if $R$ is identically zero, asymptotic flatness is already ensured. So now a theorem can be stated as 
\par 
{\it If an asymptotically flat spherical black hole solution has to be endowed with a hair (i.e. information) for a particular field, the spacetime is either Ricci-flat or the Ricci curvature falls off at least as rapidly as $\frac{1}{r^{3}}$.} 
\par As the Ricci curvature and the trace of the energy-momentum tensor are related as  $R = 8\pi G T$, the theorem can be stated in terms of the matter distribution as 
\par 
{\it If an asymptotically flat spherical black hole has a hair corresponding to a particular field, then the trace of the energy momentum tensor is either identically zero or falls off at least as fast as $\frac{1}{r^{3}}$.}
\par The theorem tells us about the necessary condition for the existence of a hair, but does not ensure anything about the sufficiency condition. It also deserves mention that the standard hair (allowed by the No Hair Conjecture) of mass and the electric charge of a spherical black hole can easily be related to the constants $C_{1}$ and $C_{2}$ respectively.
\par Very recently Faraoni and Sotiriou\cite{valerio} showed the non existence of a scalar hair in a non-minimally coupled scalar tensor theory for an axially symmetric asymptotically flat charged black black hole. It is interesting to note that their proof depends on the trace-free property of the energy momentum tensor of the associated electromagnetic field. The present work, on the other hand, deals with the trace of the field for which the hair is sought rather than that of other associated fields.

\section{Examples}
There are two most talked about counter examples of the No Hair Conjecture. The first one is the scalar hair discovered way back in 1974 by Bekenstein\cite{beken} for a nonminimally coupled conformally invariant scalar field. In the action, the Ricci scalar $R$ is coupled to the scalar field $\phi$ as $(1 - \frac{{\phi}^{2}}{6})R$. Einstein field equations are given by
\be
(1 - \frac{{\phi}^{2}}{6}) R^{\mu}_{\nu} = u^{;\alpha}_{;\alpha} {\delta}^{\mu}_{\nu} - 4u^{;\mu}_{;\nu} + 2uu^{;\mu}_{;\nu},
\ee
where $u = (1 - \frac{{\phi}^{2}}{6})$. The wave equation for the scalar field is
\be
u^{;\alpha}_{;\alpha} = 0.
\ee
These equations evidently show that the Ricci scalar $R$ and hence the trace of the energy monetum tensor $T$ are zero (see \cite{absdc}), directly verifying the result obtained in the present work.
\par The second serious counter example is that of the dilaton hair\cite{garfinkle}. The relevant metric looks like 
\be
ds^{2} = \frac{1- (2Me^{{\phi}_{0}})/r}{1 - (Q^{2}e^{3{\phi}_{0}})/(Mr)} dt^{2} -[(1-2Me^{{\phi}_{0}}/r)(1-Q^{2}e^{3{\phi}_{0}})/(Mr)]^{-1} dr^{2} -r^{2}d{\Omega}^{2}.
\ee
Here $M$, $Q$ and ${\phi}_{0}$ are the mass, the electric charge and the scalar charge respectively. This scalar charge is visible for an exterior observer. If one now calculates the Ricci scalar, it is seen that the asymptotic behaviour is dominated by $\frac{1}{r^{3}}$. This is again consistent with the theorem discussed in the present work. The counter-examples of the No Hair Conjecture are therefore found to be consistent with the theorem developed here.
\par It also deserves mention that the conclusions regarding a black hole hair remains the same for a conformal transformation of the metric\cite{saa}, so the conclusions dervied here in the string frame will be valid for a conformally tranformed frame as well.
\par  It should be good to check how do the examples favouring the No Hair Conjecture behave vis-a-vis the present theorem. We pick up one example, namely, Penney's well known solution\cite{penney} for a scalar field distribution along with an electromagnetic field.The solution is given by 
\begin{equation}
ds^{2}=e^{\gamma}dt^{2}-e^{\alpha}dr^{2}-e^{\beta}d{\Omega}^{2},
\end{equation}
where $\alpha + \gamma=0$ like the present case. The metric functions are given as
$$e^{\alpha}= (r^{2}-2mr+\frac{K{\epsilon}^{2}}{2A^{2}})^{-A} (\frac{b(r-a)^{A} -a(r-b)^{A}}{b-a})^{2},$$ and \\
$$e^{\beta}=(r^{2}-2mr+\frac{K{\epsilon}^{2}}{2A^{2}})e^{\alpha}.$$ \\
The constants are related by 
$2A^{2}ab=K{\epsilon}^{2},$ $a+b=m,$ and 
$A^{2}Kc^{2}=(1-A^{2})(2A^{2}m^{2}-K{\epsilon}^{2}).$
Here $m$ and $\epsilon$ are the mass and charge of the distribution and $c$ is the scalar charge, which is zero if $A=1.$
If one demands a non trivial scalar hair for an exterior observer out of this solution, the so called horizon becomes singular and one does not have a black hole. Thus, the solution does not yield a non-trivial scalar hair. It can be shown that the trace of the energy mementum tensor in fact falls off as $\frac{1}{{\rho}^{2}}$ where $\rho$ is the proper radial distance. So $T$ falls off slower than $\frac{1}{{\rho}^{3}}$, the minimum rate required by the theorem, and thus do not gives rise to a non trivial hair. It should be noted that the radial coordinate in Penney's work is not the proper radius, and the asymptotic behaviour of $T$ or $R$ is carefully examined against the proper radius $\rho$, given by $\rho = e^{\beta},$ and not against the radial coordinate $r.$
\par Recently Martinez and Troncoso\cite{martinez} reported the the existence of a scalar hair for a minimally coupled scalar field endowed with a potentail $V(\phi)=cosh^{4}\alpha \phi$ where $\alpha$ is a constant and $\phi$ is the scalar field. But the solution is not asymptotically flat and hence does not come in the purview of the present theorem. Incidentally, the solution is asymptotically anti-deSitter. 
\section{Discussion}
With a somewhat moderate ambition, a theorem towards characterization of the fields leading to a hair for an asymptotically flat black hole has been proved. The term ``hair'' indicates a non-trivial information regarding the corresponding field for an exterior observer. The ambition is restricted as the theorem is proved only for a spherical black hole. An axially symmetric black hole, however, would add only the information regarding the angular momentum of the black hole, which does not add to the trace of the energy momentum tensor. So the theorem is at least intuitively correct for axially symmetric black holes as well. The two most talked about counter-examples of the no hair conjecture are absolutely compatible with the present theorem. The Bekenstein black hole\cite{beken} provides an example of a primary hair and the dilaton black hole gives an example of a secondary hair which grows on the mass and the charge of the black hole. So both classes of black hole hair is included in the purview of the theorem. The theorem in fact proves a necessary condition and not a sufficient condition on the matter distribution for the existence of a hair. It is interesting to note that common examples ( eg Penney's solution\cite{penney}) in favour of the No Hair Conjecture as well as the counter-examples both are completely consistent with the theorem discussed in the present work.\\
\par Another point that deserves mention is that the present work, i.e., the theorem that has been proved, does not talk about any particular field like a scalar field or a vector field or anything, this is fairly general. Although the proof is based on fields which are minimally coupled to gravity, this in principle takes care of the nonminimally coupled theories as well. This is because by virtue of a conformal transformation, one can reduce such theories into a minimally coupled one at least formally, and it has been shown that the conclusion regarding the existence of a black hole hair is independent of this choice of frame\cite{saa}.
\par The present work depends crucially on the asymptotic flatness. There could be possibilities of asymptotically non-flat solutions in the literature. This should also be looked at. One such example is already there\cite{martinez}. The other direction of investigation will certainly be to include non-spherical black holes. As already mentioned, some work has started in that direction too\cite{valerio}.

\end{document}